\documentclass[12pt,preprint]{aastex}

\providecommand{\boldsymbol}[1]{\mbox{\boldmath $#1$}}

\providecommand{\tabularnewline}{\\}

\newcommand{\Rm}{Rm} 
\newcommand{\unit}[1]{{\rm\,#1}} 
\shorttitle{2D MRI simulations}
\shortauthors{W. Liu \emph{et al.}}

\begin{document}

\title{Simulations of Magnetorotational Instability in a Magnetized 
Couette Flow}

\author{Wei Liu}

\affil{Princeton Plasma Physics Laboratory, Princeton University, P.O. Box
451, Princeton, NJ 08543, USA }
\email{wliu@pppl.gov}

\author{Jeremy Goodman}

\affil{Princeton University Observatory, Princeton, NJ 08544, USA}

\author{Hantao Ji}
\affil{Princeton Plasma Physics Laboratory, Princeton University, P.O. Box
451, Princeton, NJ 08543, USA }

\begin{abstract}
  In preparation for an experimental study of magnetorotational
  instability (MRI) in liquid metal, we present non-ideal
  two-dimensional magnetohydrodynamic simulations of the nonlinear
  evolution of MRI in the experimental geometry. The simulations adopt
  initially uniform vertical magnetic fields, conducting radial
  boundaries, and periodic vertical boundary conditions.  No-slip
  conditions are imposed at the cylinders. Our linear growth rates
  compare well with existing local and global linear analyses.  The
  MRI saturates nonlinearly with horizontal magnetic fields comparable
  to the initial axial field.  The rate of angular momentum transport
  increases modestly but significantly over the initial state.  For
  modest fluid and magnetic Reynolds numbers $Re,\Rm\sim 10^2-10^3$,
  the final state is laminar reduced mean shear 
  except near the radial boundaries, and with poloidal circulation scaling as
  the square root of resistivity, in partial agreement with the
  analysis of Knobloch and Julien.  A sequence of simulations at
  $\Rm=20$ and $10^2\lesssim Re\lesssim 10^{4.4}$ enables extrapolation to
  the experimental regime ($\Rm\approx 20$, $Re\sim 10^7$), albeit
  with unrealistic boundary conditions.  MRI should increase the
  experimentally measured torque substantially over its initial purely
  hydrodynamic value.
\end{abstract}

\maketitle

\section{Introduction}\label{sec:1}

Rapid angular momentum transport in accretion disks has been a
longstanding astrophysical puzzle. The molecular viscosity of
astrophysical gases and plasmas is completely inadequate to explain
observationally inferred accretion rates, so that a turbulent
viscosity is required.  Recent theoretical work
\citep{pj81,bh91,hb91,bh98} indicates that purely hydrodynamic
instabilities are absent or ineffective, but that magnetorotational
instabilities (MRI) are robust and support vigorous turbulence in
electrically-conducting disks. Although originally discovered by
\citet{ve59} and \citet{chan60}, MRI did not come to the attention of the
astrophysical community until rediscovered by \citet{bh91} and
verified numerically \citep{hgb95,bnst95,mt95}. It is now believed
that MRI drives accretion in disks ranging from quasars and X-ray
binaries to cataclysmic variables and perhaps even protoplanetary
disks \citep{bh98}. Some astrophysicists, however, argue from
laboratory evidence that purely hydrodynamic turbulence may account
for observed accretion rates, especially in cool, poorly
conducting disks where MRI may not operate
\citep{db93,rz99,dsb00,hrz01}.

\begin{figure}[!htp]
\caption{\label{fig:cylinder}Geometry of Taylor-Couette
flow.  In the Princeton MRI experiment, $r_1=7.1\unit{cm}$,
$r_2=20.3\unit{cm}$, $h=27.9\unit{cm}$.}
\epsscale{0.40}
\plotone{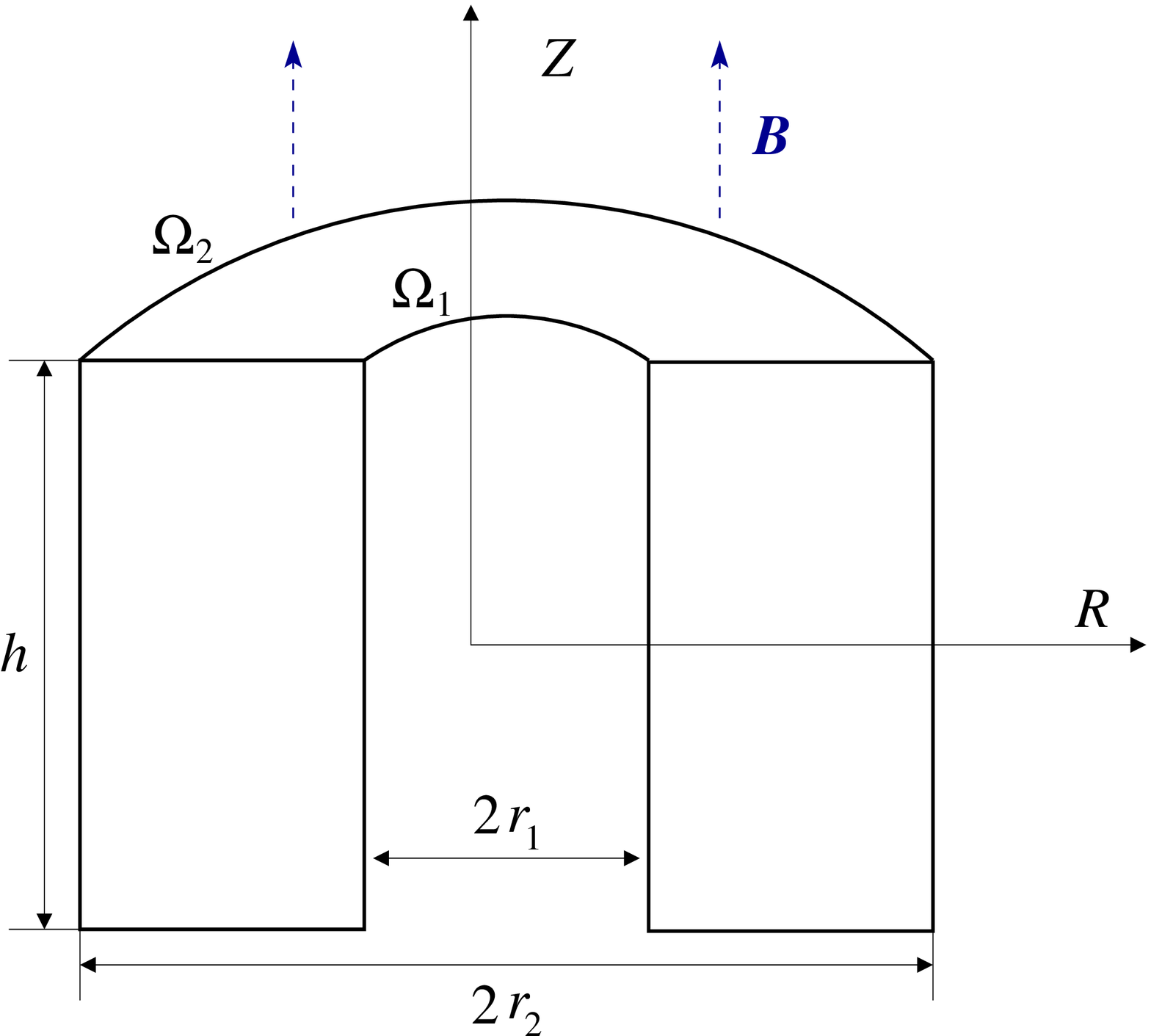}

\end{figure}

Although its existence and importance are now accepted by most
astrophysicists, MRI has yet to be clearly demonstrated in the
laboratory, notwithstanding the claims of \citet{sisan04}, whose
experiment proceeded from a background state that was not in MHD
equilibrium.  Recently\citep{jgk01,gj02}, we have therefore proposed
an experimental study of MRI using a magnetized Couette flow: that is,
a conducting liquid (gallium) bounded by concentric differentially
rotating cylinders and subject to an axial magnetic field.
The radii of the cylinders are
$r_{1}<r_{2}$, as shown in Fig.~\ref{fig:cylinder};
their angular velocities, $\Omega_{1}$ \& $\Omega_{2}$, have
the same sign in all cases of interest to us.
If the cylinders were infinitely
long---very easy to assume theoretically, but rather more difficult to
build experimentally---the steady-state solution would be ideal
Taylor-Couette flow:
\begin{equation}
\Omega(r)=a+\frac{b}{r^{2}}
\end{equation}
where $a=(\Omega_{2}r_{2}^{2}-\Omega_{1}r_{1}^{2})/(r_{2}^{2}-r_{1}^{2})$
and $b=r_{1}^{2}r_{2}^{2}(\Omega_{1}-\Omega_{2})/(r_{2}^{2}-r_{1}^{2})$.
In the unmagnetized and inviscid limit, such a flow is
linearly axisymmetric stable if and only if the specific
angular momentum increases outwards: that is,
$(\Omega_{1}r_1^2)^2<(\Omega_{2}r_2^2)^2$, or equivalently, $ab>0$.
A vertical magnetic field may destabilize the flow, however,
provided that the angular \emph{velocity} decreases outward,
$\Omega_{2}^2<\Omega_{1}^2$; in ideal MHD, instability
occurs at arbitrarily weak field strengths \citep{bh91}.
The challenge for experiment, however, is that liquid-metal flows are
very far from ideal on laboratory scales.  While the fluid Reynolds
number $Re\equiv \Omega_{1}r_{1}(r_{2}-r_{1})/\nu$
can be large, the corresponding \emph{magnetic} Reynolds number
\begin{equation}\label{eq:Re}
\Rm\equiv\frac{\Omega_{1}r_{1}(r_{2}-r_{1})}{\eta}
\end{equation}
is modest or small, because the magnetic Prandtl number 
$P_{\rm m}\equiv\nu/\eta\sim 10^{-6}$ in liquid metals.
Standard MRI modes will not grow unless both the rotation period
and the Alfv\'en crossing time are shorter than the timescale
for magnetic diffusion.  This requires both $\Rm\gtrsim 1$ and
$S\gtrsim 1$, where
\begin{equation}
S\equiv\frac{V_{A}(r_{2}-r_{1})}{\eta}
\end{equation}
is the Lundquist number, and
$V_{A}=B/\sqrt{4\pi\rho}$ is the Alfv\'en speed.
Therefore, $Re\gtrsim 10^6$ and fields of several kilogauss
must be achieved in typical experimental geometries.

Recently, it has been discovered that MRI modes may grow at much
reduced $\Rm$ and $S$ in the presence of a helical background field, a
current-free combination of axial and toroidal
field \citep{hr05,rhss05}.  We have investigated these helical MRI
modes.  While we confirm the quantitative results given by the
authors just cited for the onset of instability, we have uncovered
other properties of the new modes that cast doubt upon both their
experimental realizability and their relevance to astrophysical disks.
To limit the length of the present paper, we present results for
purely axial background fields only.  A paper on helical MRI is in
preparation.

One may question the relevance of experimental to
astrophysical MRI, especially its nonlinear phases.  In accretion
disks, differential rotation arises from radial force balance between
the gravitational attraction of the accreting body and centrifugal
force.  Thermal and magnetic energies are small compared to orbital
energies, at least if the disk is vertically thin compared to its
radius.  Consequently, nonlinear saturation of MRI cannot occur by
large-scale changes in rotation profile.  In experiments, however,
differential rotation is imposed by viscous or other weak forces, and
the incompressiblity of the fluid and its confinement by a container
allow radial force balance for arbitrary $\Omega(r)$.  Thus,
saturation may occur by reduction of differential rotation, which is
the source of free energy for the instability.  In this respect, MRI
experiments and the simulations of this paper may have closer
astrophysical counterparts among differentially rotating stars, where
rotation is subsonic and boundaries are nearly stress-free 
\citep{bh94,mbs05}.
Both in the laboratory and in
astrophysics, however, nonlinear MRI is expected to enhance the radial
transport of angular momentum. Quantifying the enhanced transport in
a Couette flow is a primary goal of the Princeton MRI experiment
and of the present paper.

Another stated goal of the Princeton experiment is to validate
astrophysical MHD codes in a laboratory setting.  Probably the most
widely used astrophysical MHD code is ZEUS \citep{sn921,sn922},
which exists in several variants.  The simulations of this paper use
ZEUS-2D.  Like most other astrophysical MHD codes, ZEUS-2D was
designed for compressible, ideal-MHD flow with simple boundary
conditions: outflow, inflow, reflecting---but not no-slip.  ZEUS would
not be the natural choice of a computational fluid-dynamicist
interested in Couette flow for its own sake.  Nevertheless, after
modifying ZEUS-2D to incorporate resistivity, viscosity, and no-slip
boundary conditions, we find it to be a robust and flexible tool for
the subsonic flows of interest to us.  It reproduces the growth rates
predicted for incompressible flow (\S\ref{sec:3}), and agrees with
hydrodynamic laboratory data \citep{brj05}; MHD data are not yet
available.  Of course, all real flows are actually compressible; in
an ideal gas of fixed total volume, density changes generally scale
$\sim M^2$ when Mach number $V_{\rm flow}/V_{\rm sound}< 1$.
Incompressibility is an idealization in the limit $M\to0$.  We have
used an isothermal equation of state in ZEUS with a sound speed chosen
so that the maximum of $M\le 1/4$ and obtain quantitative agreement
with incompressible codes at the few-percent level (\S\ref{sec:2}).

Most of the parameters of the simulations in \S\S\ref{sec:3}-\ref{sec:4} 
are chosen to match those of the experiment.
We adopt the same cylinder radii (Fig.~\ref{fig:cylinder}).
The experimental rotation rates of both cylinders (and of the endcaps)
are separately adjustable, as is the axial magnetic field.  For these
simulations, we adopt fixed values within the achievable range:
$\Omega_{1}=4000\unit{rpm}$ \& $\Omega_{2}=533\unit{rpm}$,
$B_{z0}=5000\unit{G}$.  We set the density of the fluid to that
of gallium, $\rho=6\unit{g\,cm^{-3}}$.

Our simulations depart from experimental reality in two important
respects: Reynolds number and vertical boundary conditions.
Computations at $Re\gtrsim 10^6$ are out of reach of any present-day
code and computer, at least in three dimensions; $Re\sim 10^6$ might
just be achievable in axisymmetry, but higher-$Re$ flows are more
likely to be three-dimensional, so that an axisymmetric simulation at
such a large $Re$ is of doubtful relevance.  (The same objection might
be leveled at all of our simulations for $Re\gg 10^3$.  Those
simulations are nevertheless useful for establishing scaling
relations, even if the applicability of the relations to real
three-dimensional flows is open to question.)  We use an artificially
large kinematic viscosity so that $Re=10^2-10^{4.4}$, whereas for the true
kinematic viscosity of gallium ($\nu \simeq
3\times10^{-3}\unit{cm^{2}s^{-1}}$), $Re\approx 10^7$ at the
dimensions and rotation rates cited above.  In defense of this
approximation, we point to the fact that extrapolations of
Ekman-circulation rates and rotation profiles simulated at $Re<10^4$
agree well with measurements taken at $Re=10^6$ both in a prototype
experiment \citep{kjg04}, and in the present aparatus \citep{brj05}.
We \emph{are} able to reproduce the experimental values of the
dimensionless parameters based on resistivity: $\Rm\sim 20$, $S\sim
4$; we also report simulations at $\Rm\sim 10^2-10^4$.  (The actual
diffusivity of gallium is $\eta \simeq
2\times10^{3}\unit{cm^{2}s^{-1}}$).

Except for hydrodynamic test simulations carried out to compare with
incompressible results and laboratory data (\S\ref{sec:2}), we adopt
vertically periodic boundary conditions for all fluid variables, with
a periodicity length $L_z=2h$, where $h=27.9 \unit{cm}$ is the actual height of
the experimental flow.  Such boundary conditions are physically
unrealistic, but almost all published linear analyses of MRI in Couette flows
have adopted them because they permit a complete separation of
variables \citep{jgk01,gj02,npc02,rs02,rss03}; an exception
is \citet{rz01}. Thus by adopting
periodic vertical boundaries, we are able to test our code against
well-established linear growth rates and to explore---apparently for
the first time in Couette geometry---the transition from linear growth
to nonlinear saturation.  The imposition of no-slip conditions at finite
endcaps introduces important complications to the basic state, including
Ekman circulation and Stewartson layers, which we are currently studying,
especially as regards their modification by the axial magnetic field.
But the experimental apparatus has been designed to minimize these
complications (\emph{e.g.} by the use of independently controlled split
endcaps) in order to approximate the idealized Couette flows
presented here, whose nonlinear development already presents features
of interest.  This paper is the first in a series; later papers will
address the effects of finite endcaps on magnetized flow, helical MRI
instabilities, \emph{etc.}


\section{Modifications to ZEUS-2D and Code Tests}\label{sec:2}

ZEUS-2D offers the option of cartesian $(x,y)$, spherical
$(R,\theta)$, or cylindrical $(z,r)$ coordinates.  We use $(z,r)$.
Although all quantities are assumed independent of the azimuth
$\varphi$, the azimuthal components of velocity ($v_\varphi$) and
magnetic field ($B_\varphi$) are represented.  We have implemented
vertically periodic boundary conditions (period$=2h$) for all
variables, and conducting radial boundary conditions for
the magnetic field.  Impenetrable, no-slip radial boundaries are
imposed on the velocities.  Viscosity and resistivity have been added
to the code.  In order to conserve angular momentum precisely, we cast
the azimuthal component of the Navier-Stokes equation in conservative
form:
\begin{equation}\label{aziNS}
\frac{\partial L}{\partial t} ~+~\frac{\partial}{\partial z}
\left(V_z L + F_z\right)~+~
\frac{1}{r}\frac{\partial}{\partial r}\left( r V_r L + r F_r\right) =0,
\end{equation}
in which $L=r V_\varphi$, and $F_r$ and $F_z$ are the viscous
angular-momentum fluxes per unit mass,
\begin{equation}
F_z= -\nu\frac{\partial L}{\partial z}\,,\qquad
F_r= -\nu r^2\frac{\partial}{\partial r}\left(\frac{L}{r^2}\right)\,.
\end{equation}
In the spirit of ZEUS, the viscous part of eq.~(\ref{aziNS}) is
implemented as part of the ``source'' substep.
In accord with
the Constrained Transport algorithm \citep{eh88}, which preserves
$\nabla\cdot\mathbf{B}=0$, resistivity is implemented by
an ohmic term added to the electromotive force, which becomes
\begin{equation}\label{emf}
\mathbf{\mathcal{E}}=\mathbf{V}\times\mathbf {B}
-\eta \mathbf{\nabla\times B}\,.
\end{equation}

\subsection{Code Tests (1) - Wendl's Low-$Re$ Solution}

At $Re\ll 1$ and $\Rm=0$, poloidal flow is negligible and the
toroidal flow is steady.  $V_\varphi$ satisfies
\begin{equation}
\nu(\bigtriangledown^{2}-\frac{1}{r^{2}})V_{\varphi}=0.
\end{equation}
\citet{wm99} has given the analytic solution
of this equation for no-slip vertical boundaries co-rotating with
the outer cylinder. This serves as one benchmark for the viscous part
of our code; note that the vertical boundary conditions differ from those
used in the simulations of \S\ref{sec:3}-\ref{sec:4}.

\begin{figure}[!htp]
\caption{\label{Fi:2} Radial profile of the azimuthal velocity for $Re=1$.}
\epsscale{0.40}
\plotone{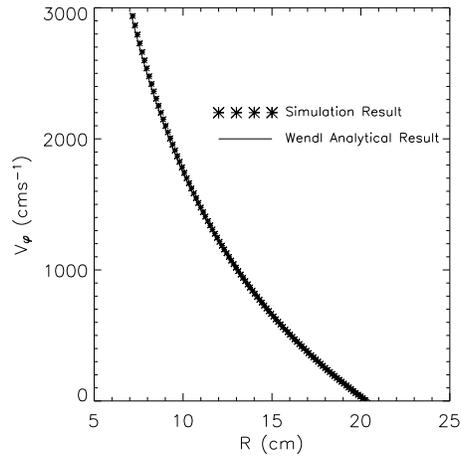}
\end{figure}

Figure~\ref{Fi:2} compares results from ZEUS-2D with the analytical result.
The maximum relative error is less than $3\%$.
We have also calculated the viscous torque across the mean
cylinder $(r=(r_{1}+r_{2})/2)$. 
Wendl's solution predicts $-1.5004\times 10^{9}\unit{g\,cm^{2}\,s^{-2}}$, 
and our simulations yield $-1.5028\times 10^{9}\unit{g\,cm^{2}\,s^{-2}}$.

\subsection{Code Tests (2) - Magnetic Diffusion}

If the fluid is constrained to be at rest, then the toroidal
induction equation becomes
\begin{equation}
\frac{\partial B_{\varphi}}{\partial t}=\eta
\left(\frac{\partial^{2}B_{\varphi}}{\partial r^{2}}+
\frac{1}{r}\frac{\partial B_{\varphi}}{\partial r}-
\frac{B_{\varphi}}{r^{2}}+\frac{\partial^{2}B_{\varphi}}{\partial z^{2}}\right)
\end{equation}
An exact solution compatible with our boundary conditions is:
\begin{equation}
\mathbf{B}=\mathbf{\hat e}_{z}B_{z}^{0}
~+\mathbf{\hat e}_\varphi\frac{B_{\varphi}^{0}}{r}\cos(kz)
\exp(-\eta k^{2}t)
\end{equation}where $k$ is the wave number, and
$B_z^0$ and $B_r^0$ are constants.

A comparison of the theoretical and simulated results shows that the
error scales quadratically with cell size, as expected for our
second-order difference
scheme (Table~\ref{cap:Gauss-Magnetic-Diffusion}).

\subsection{Comparison with an Incompressible Code}

\begin{figure}[!htp]

\caption{\label{cap:Comparison-with-incompressible}Comparison with
incompressible code at $Re=1600$ : (a) Contours of toroidal velocity
from \citet{kjg04} (b) Results from ZEUS-2D with $M=1/4$ }
\epsscale{1.00} 
\plotone{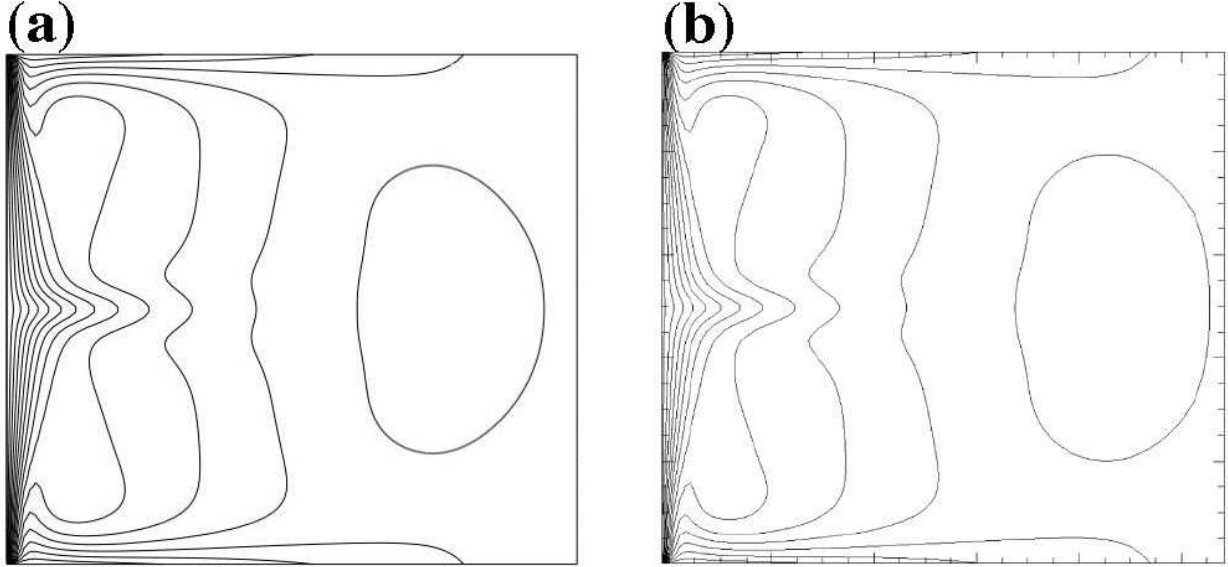}
\end{figure}

ZEUS-2D is a compressible code. However our experimental fluid,
gallium, is nearly incompressible at flow speeds of interest, which
are much less than its sound speed, $2.7\unit{km\,s^{-1}}$.  As
mentioned in \S1, we can approximate incompressible flow by using a
subsonic Mach number, $M<1$.  However, since ZEUS is explicit, $M\ll1$
requires a very small time step to satisfy the CFL stability
criterion.  As a compromise, we have used $M=1/4$ (based on the inner
cylinder) throughout all the simulations presented in this paper.
We assume an isothermal equation of state to avoid increases in $M$
by viscous and resistive heating; the nonlinear compressibility and
thermodynamic properties of the actual liquid are in any case very
different from those of ideal gases, for which ZEUS was written.
Figure~\ref{cap:Comparison-with-incompressible} compares results
obtained from ZEUS-2D with simulations performed by \citet{kjg04} using
their incompressible Navier-Stokes code.

\section{Linear MRI Simulations}\label{sec:3}

In the linear regime, MRI has been extensively studied both locally
and globally \citep{jgk01,gj02,rz01,npc02,rs02,rss03}. We have used
these linear results to benchmark our code.

In the linear analyses cited above, the system is assumed to be
vertically periodic with periodicity length $2h$, twice the height
height of the cylinders.  In cylindrical coordinates, the equilibrium
states are $\mathbf{B_{0}}=B_0\mathbf{\hat e}_z$ and
$\mathbf{V_{0}}=r\Omega\mathbf{\hat e}_\varphi$.  WKB methods describe
the stability of this system very well even on the largest scales
\citep{jgk01,gj02}.  Linear modes are proportional to $\exp(\gamma
t-ik_{z}z)f(k_{r}r)$, where $\gamma$ is the growth rate, and $f(x)$ is
an approximately sinusoidal radial function, at least outside boundary
layers, whose zeros are spaced by $\Delta x\approx\pi$.  The
wavenumbers $k_{z}=n\pi/h$ and $k_{r}\approx m\pi/(r_2-r_1)$, where
$n$ and $m$ are positive integers.  We will consider only the lowest
value of $k_r$ ($m=1$) but allow $n\ge1$.  The initial perturbation is
set to an approximate eigenmode appropriate for conducting boundary
conditions:
\begin{eqnarray}\label{initmodes}
\delta B_{z}&=&A\sin k_{z}z\frac{r_{1}+r_{2}-2r}{r}\qquad
\delta B_{r}=k_{z}A\cos k_{z}z\frac{(r_{2}-r)(r-r_{1})}{r}\qquad
\delta B_{\varphi}=0\nonumber\\[1ex]
\delta V_{z}&=&B\cos k_{z}z\frac{r_{1}+r_{2}-2r}{r}\qquad
\delta V_{r}=k_{z}B\sin k_{z}z\frac{(r_{2}-r)(r-r_{1})}{r}\qquad
\delta V_{\varphi}=0.
\end{eqnarray}

\begin{figure}[!htp]
\caption{\label{cap:Comparison-with-Local}
MRI growth rate versus $\Rm$ for conducting radial boundaries.
{\it Points:} simulations.  {\it Curve:} global linear analysis
\citep{gj02} with $Re=25,600$.}
\epsscale{0.40} \plotone{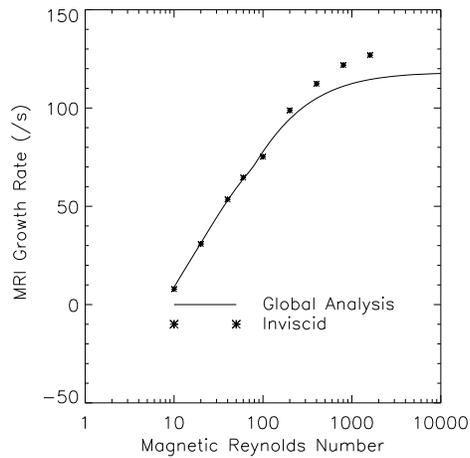}
\end{figure}

Evidently, the fast-growing mode dominates the simulations
no matter which $n$ is used initially.
Figure~\ref{cap:Comparison-with-Local} compares the MRI growth rate
obtained from the simulations with those predicted by
global linear analysis
\citep{gj02} as a function of magnetic Reynolds number.

The radially global, vertically periodic linear analysis of
\citet{gj02} found that the linear eigenmodes have boundary layers
that are sensitive to the dissipation coefficients, but that the
growth rates agree reasonably well with WKB estimates except near
marginal stability.  A comparison of the growth rates found by this
analysis with those obtained from our simulations is given in
Table~\ref{cap:Comparison-with-Global}.  In the context of the
simulations, ``$Re=\infty$'' means that the explicit viscosity
parameter of the code was set to zero, but this does not guarantee
inviscid behavior since there is generally some diffusion of angular
momentum caused by finite grid resolution.  Nevertheless, since the
magnetic Reynolds number of the experiment will be about $20$ and
since $Re/\Rm\sim 10^6$, these entries of the table probably most
closely approximate the degree of dissipation in the gallium
experiment.  In Table~\ref{cap:Comparison-with-Global}, the largest
growth rate predicted by the linear analysis has been marked with an
asterisk ({*}).  The simulations naturally tend to be dominated by the
fastest numerical mode---that is, the fastest eigenmode of the
finite-difference equations, which need not map smoothly into the
continuum limit.  Fortunately, as asserted by the Table, the fastest
growth occurs at the same vertical harmonic $n$ in the simulations as
in the linear analysis.

\section{Nonlinear Saturation}\label{sec:4}

\begin{figure}[!htp]

\caption{\label{cap:Rotating-Speed-Profile} Angular velocity profile before
and after saturation at several heights, for 
$Re=\Rm=400$. ``Jet'' is centered at $z=0$ (squares).}
\epsscale{0.60} \plotone{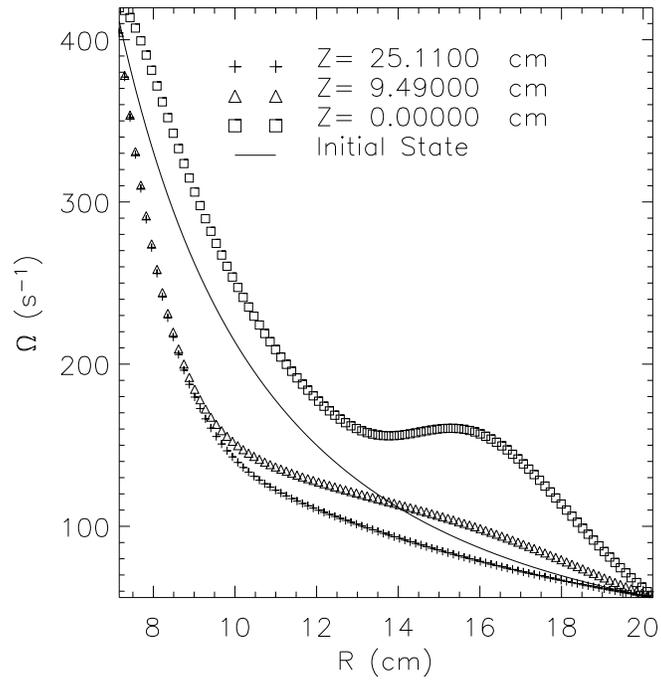}

\end{figure}

As noted in \S\ref{sec:1}, instabilities cannot easily modify the
differential rotation of accretion disks because internal and magnetic
energies are small compared to gravitational ones, and MRI is believed
to saturate by turbulent reconnection \citep{fsh00,si01}.  In Couette
flow, however, the energetics do not preclude large changes in the
rotation profile.  As shown by
Fig.~\ref{cap:Rotating-Speed-Profile}), the differential rotation of
the final state is reduced somewhat compared to the initial state in
the interior of the flow, and steepened near the inner cylinder.

\subsection{Structure of the final state}

\begin{figure}[!htp]

\caption{\label{cap:Flow-Patterns-of} Contour plots of final-state
velocities and fields.  $Re=400$, $\Rm=400$. (a) Poloidal flux
function $\Psi \unit{(Gauss\,cm^{2})}$ (b) Poloidal stream function
$\Phi \unit{(cm^{2}s^{-1})}$ (c) toroidal field
$B_{\varphi} \unit{(Gauss)}$ (d) angular velocity
$\Omega\equiv r^{-1}V_{\varphi}\unit{(rad\,s^{-1})}$} 
\epsscale{1.00}
\plotone{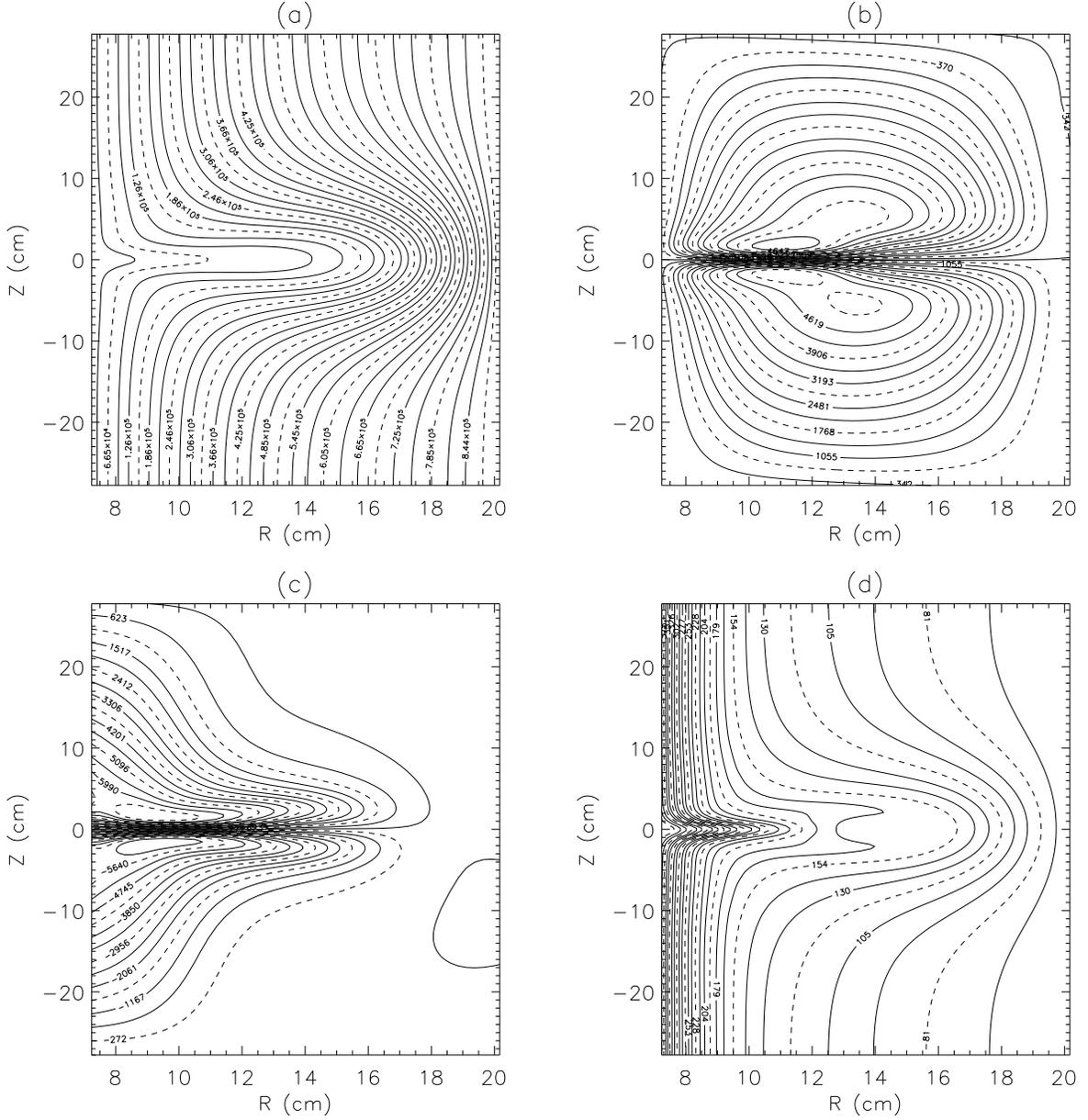}

\end{figure}

For moderate dissipation ($Re,\Rm\lesssim 10^3$), 
the final state is steady.  Typical flow and
field patterns are shown in
Figure~\ref{cap:Flow-Patterns-of}.  The poloidal flux
and stream functions are defined so that
\begin{equation}\label{pfuncs}
\boldsymbol{V}_P\equiv
V_r\boldsymbol{e}_r+V_z\boldsymbol{e}_z=r^{-1}\boldsymbol{e}_\varphi
\boldsymbol{\times\nabla}\Phi,\qquad \boldsymbol{B}_P\equiv
B_r\boldsymbol{e}_r+B_z\boldsymbol{e}_z=r^{-1}\boldsymbol{e}_\varphi
\boldsymbol{\times\nabla}\Psi,
\end{equation}
which imply $\boldsymbol{\nabla\cdot V}_P=0$
and $\boldsymbol{\nabla\cdot B}_P=0$.
[Our velocity field is slightly compressible, so that eq.~(\ref{pfuncs})
does not quite capture the full velocity field.
Nevertheless, the error is small, and 
$\Phi$ is well defined by
$\nabla^2(\Phi\boldsymbol{e}_\varphi/r)=
\boldsymbol{\nabla\times V}_{\rm P}$ with periodic
boundary conditions in $z$ and $\partial\Phi/\partial z=0$ on the cylinders.]

\begin{figure}[!htp]

\caption{\label{cap:jet} Radial velocity versus $z$ for $Re=400$, at
several radii (cm): $+,~8.42$; $*,~10.27$; $\times,~11.98$;
{\small$\triangle$},~$13.70$; $\Diamond,~16.87$; $\Box,~18.98$.  For
clarity, only half the full vertical period ($56\unit{cm}$) is shown.
Panel (a), $\Rm=400$; panel (b) $\Rm=6400$.}
\epsscale{1.00}
\plotone{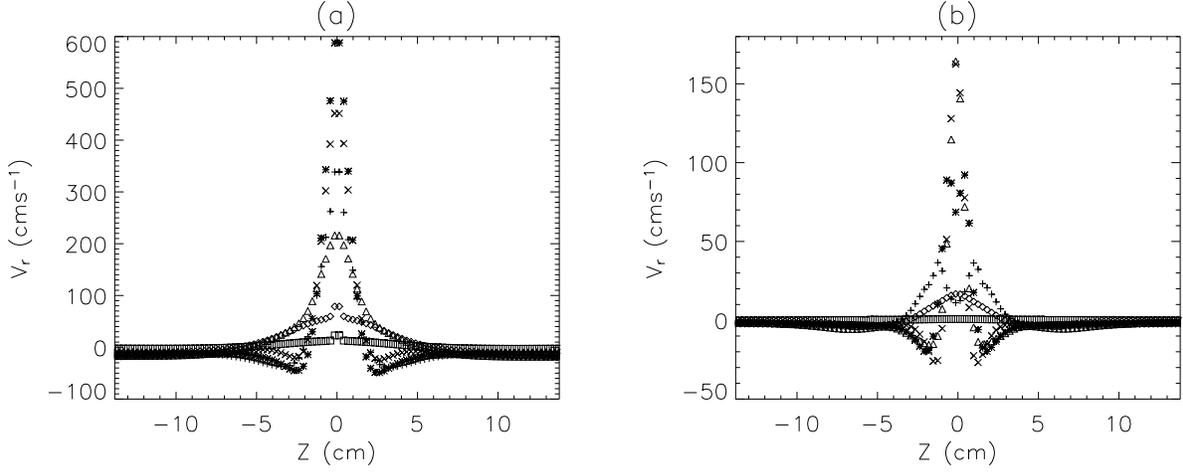}

\end{figure}

The most striking
feature is the outflowing ``jet'' centered near $z=0$
in Figure~\ref{cap:Flow-Patterns-of}.
The contrast in flow speed between the
jet and its surroundings is shown more clearly in
Figure~\ref{cap:jet}. Figure~\ref{cap:Flow-Patterns-of}
also shows that the horizontal
magnetic field changes rapidly across the jet, which therefore
approximates a current sheet.

\begin{figure}[!htp]
\caption{\label{cap:jetfit} Maximum radial speed in the jet (left panel) and
maximum of poloidal stream function (right panel) \emph{vs.} magnetic Reynolds number,
for $Re=400$.  Powerlaw fits are shown as dashed lines with slopes $-0.53$ [left panel, eq.~(\ref{jetspeed})]
and $-0.57$ [right panel].}
\epsscale{1.00}
\plotone{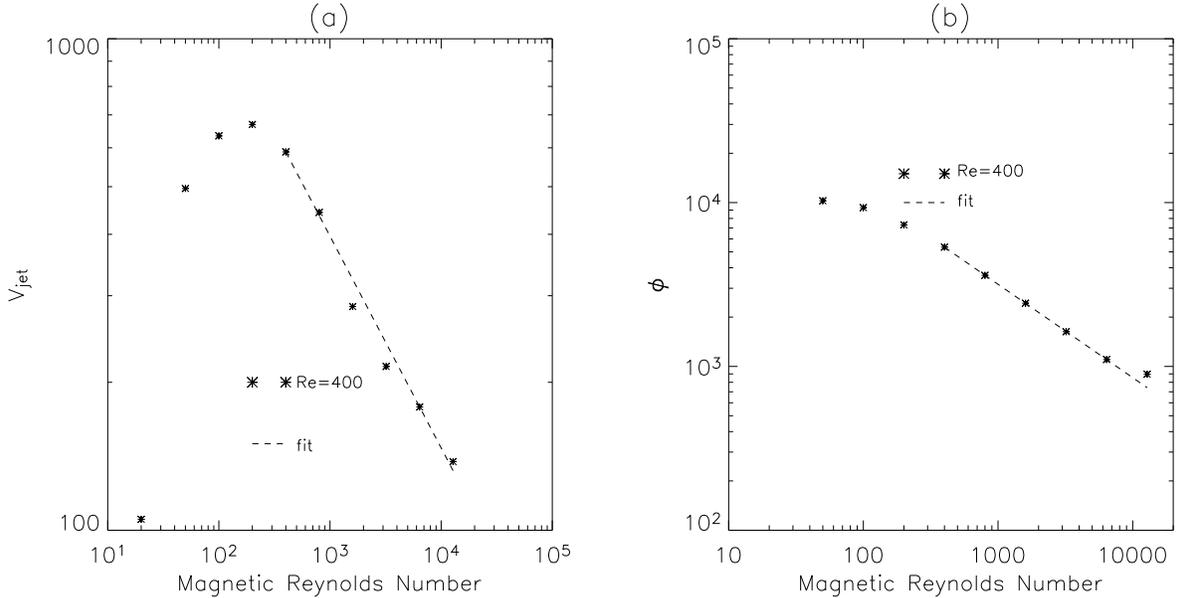}
\end{figure}

The radial flow speed in the jet scales with $\Rm$ as 
(Fig.~\ref{cap:jetfit}), 
\begin{equation}\label{jetspeed}
V_{\rm jet}\propto \Rm^{-0.53}.
\end{equation}
We find that the radial speed outside the jet scales similarly,
\begin{equation}\label{externalspeed}
V_{\rm external}\propto \Rm^{-0.56} \propto \eta^{0.56}.
\end{equation}
Mass conservation demands that
$V_{\rm jet}W_{\rm jet}=V_{\rm external}(2 h-W_{\rm jet})$,
where $W_{\rm jet}$ is the effective width of the jet. 
Thus we can conclude that
this width is independent of magnetic Reynolds  number:
\begin{equation}\label{cap:jetwidth}
W_{\rm jet}\propto \Rm^{0}
\end{equation}
Additional support for this conclusion comes from the nearly equal scaling
of $V_{r}$ and $\Phi$ with $\Rm$ (Fig.~\ref{cap:jetfit}), which
indicates that the spatial scales in the velocity field are 
asymptotically independent of $\Rm$.
The toroidal flow perturbation and toroidal field are
comparable to the rotation speed and initial background field, respectively:
\begin{equation}\label{ratios}
1.18 \lesssim \max\,\frac{B_{\varphi}}{B_{z0}} \lesssim 1.52, \qquad
0.28 \lesssim \max\,\frac{\delta V_{\varphi}}{r_1\Omega_1} \lesssim 0.56
\end{equation}

We emphasize that the scalings (\ref{jetspeed})-(\ref{ratios})
have been established for a limited range of flow parameters,
$10^2\lesssim Re,\Rm\lesssim 10^{4.4}$.  The jet is less well defined at
lower $\Rm$, especially in the magnetic field.  Extrapolation of these
scalings to laboratory Reynolds numbers ($Re\gtrsim 10^6$) is risky,
and indeed our simulations suggest that the final states are unsteady
at high $Re$ and/or high $\Rm$ (Fig.~\ref{approach}).

\subsection{Angular Momentum Transport}

\begin{figure}[!htp]

\caption{\label{cap:Z-Average-Torques-versus}$z$-integrated torques versus 
$r$.
$Re=400$, $\Rm=400$. Left panel: initial state; right: final state}
\epsscale{1.00}
\plotone{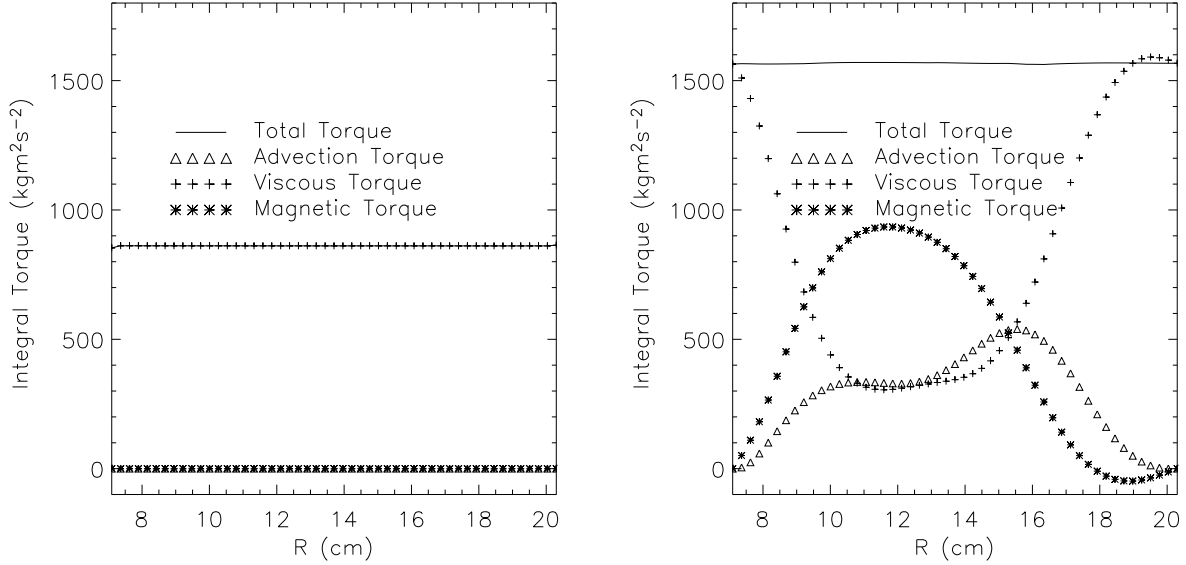}

\end{figure}

Figure~\ref{cap:Z-Average-Torques-versus} displays
the radial profiles of the advective, viscous, and magnetic
torques integrated over cylinders coaxial with the boundaries:
\begin{eqnarray}
\Gamma_{\rm advective}(r)&=&\int_{-h}^{h} dz\,\rho r^{2} v_{r}v_{\varphi}\\
\Gamma_{\rm magnetic}(r)&=&
\int_{-h}^{h} dz \left(-\frac{r^{2}B_{r}B_{\varphi}}{4\pi}\right)\\
\Gamma_{\rm viscous}(r)&=&\int_{-h}^{h} dz 
\left[-r^{3}\rho \nu \frac{\partial}{\partial r}
\left(\frac{v_{\varphi}}{r}\right)\right]\\[1ex]
\Gamma_{\rm total}(r)&=&
\Gamma_{\rm advective}(r)+\Gamma_{\rm magnetic}(r)+\Gamma_{\rm viscous}(r)
\end{eqnarray}

\begin{figure}[!htp]
\caption{Increase of total torque versus (a) $\Rm$ and (b) $Re$.
In panel (b), dashed lines have slopes of $0.5$ ($\Rm=400$) and
$0.675$ ($\Rm=20$).
}
\label{ratio}
\epsscale{1.00}
\plotone{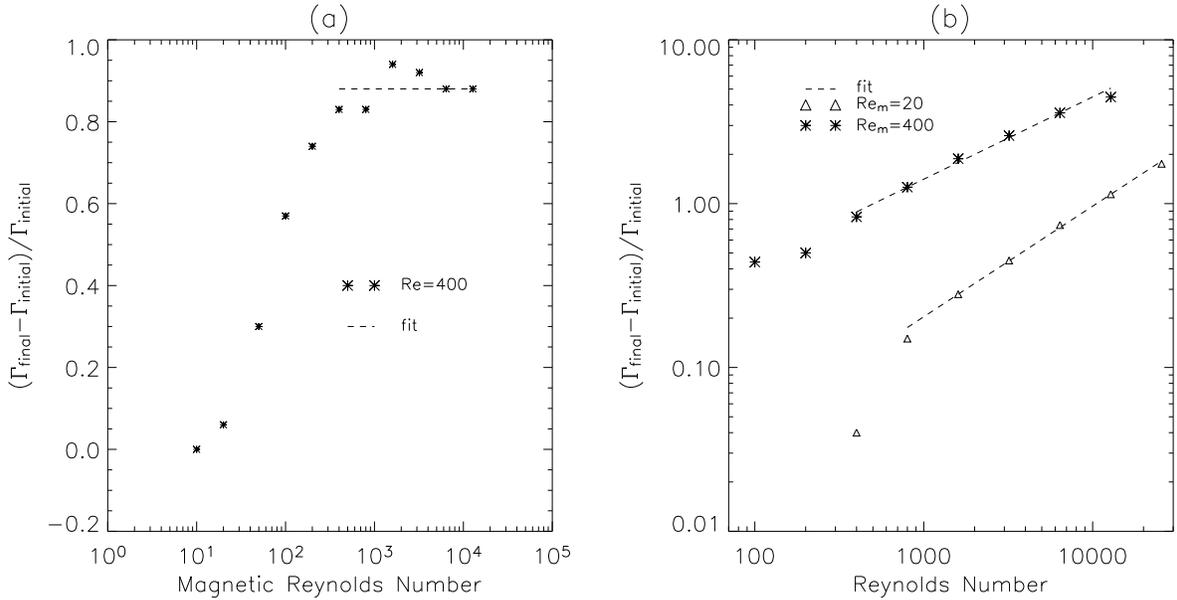}

\end{figure}

The advective and magnetic torques vanish at $r_1$ and $r_2$ because
of the boundary conditions but are important at intermediate radii. 
All components of the torque are positive except near $r_{2}$.
The total torque is constant with radius, as required in
steady state, but increases from  the initial to the final state
(Figure~\ref{cap:Z-Average-Torques-versus}).
From Figure~\ref{ratio}, we infer the scalings
\begin{equation}\label{torque_scaling}
\frac{\Gamma_{\rm final}-\Gamma_{\rm initial}}{\Gamma_{\rm initial}}
\propto Re^{0.5}\Rm^0,
\end{equation}
at least at $Re,\,\Rm\gtrsim10^3$.  In fact, a better fit to the
exponent of $Re$ for $\Rm=20$ and $Re\gtrsim 10^3$ would be $0.68$
rather than $0.5$, but the exponent seems to decrease at the largest
$Re$, and it is $\approx 0.5$ for $\Rm=400$, so we take the latter to
be the correct asymptotic value. 

Representative runs are listed in Table~\ref{tab:torques}.
Additional runs have been carried out on coarser grids (smaller $N_r,N_z$) to check
that the values quoted for the torques are independent of spatial resolution to at
least two significant figures in the laminar cases ($Re,\Rm\lesssim 10^3$) and to
better than $10\%$ in the unsteady cases where precise averages are difficult to
obtain.  In the latter cases, the quoted values in the last two columns have been
averaged over radius but not over time.

\subsection{Interpretation of the final state}

The division of the flow into a narrow outflowing jet and a slower
reflux resembles that found by \citet{kjg04} in their hydrodynamic
simulations [Fig.~\ref{cap:Comparison-with-incompressible}].  In that
case, the jet bordered two Ekman cells driven by the top and bottom
endcaps.  In the present case, however, Ekman circulation is not
expected since the vertical boundaries are periodic, and we must look
elsewhere for an explanation of the final state.

\citet[hereafter KJ]{kj05} have proposed that axisymmetric MRI may
saturate in a laminar flow whose properties depend upon the
dissipation coefficients $\nu$ \& $\eta$, with a large change in the
mean rotation profile, $\Omega(r)$.  Although this mechanism of
saturation probably cannot apply to thin disks for
the reasons given in \S\ref{sec:1}, it is consistent with
some aspects of the
final state of our Couette-flow simulations: in particular, the
scalings (\ref{jetspeed})-(\ref{externalspeed}) of the poloidal
velocities with $\Rm$; and the mean rotation profile does indeed
undergo a large reduction in its mean shear,
except near the boundaries (Fig.~\ref{cap:Rotating-Speed-Profile}).

One prominent difference between the final states envisaged by KJ and
those found here is the axial lengthscale.  KJ assumed the final state
to have the same periodicity as the fastest-growing linear MRI mode,
although they acknowledged that their theory does not require this.
In our case, the linear and nonlinear lengthscales differ: whereas the
fastest linear mode has three wavelengths over the length of the
simulation (Table~\ref{cap:Comparison-with-Global}), the nonlinear
state adopts the longest available periodicity length, namely that
which is imposed by the vertical boundary conditions.
Within that length, the flow is divided between the narrow jet and
broad reflux regions.  As discussed below, a third and even narrower
reconnection region, whose width scales differently in $\Rm$ from
that of the jet itself, exists within the jet.
Another possibly important difference concerns
the role of radial boundaries. KJ simply ignored these, yet our jet
clearly originates at the inner cylinder
(Fig.~\ref{cap:Flow-Patterns-of}).  KJ's
theory predicts that the poloidal flow should be proportional to
$Re^{-1/2}$ as well as $\Rm^{-1/2}$
Yet, we find that $V_{r,\rm jet}$ actually
\emph{increases} with $Re$, roughly as $Re^{+1/2}$, up to $Re\sim
10^3$, above which it begins to decline and the
flow becomes unsteady.

The jet is probably the part of the flow that corresponds most closely
to the ``fingers'' envisaged by KJ.  Let us at least try to understand
how the quantities in our jet scale with increasing $\Rm$ at fixed $Re$,
even though it is more relevant to the experiment to
increase $Re$ at fixed and modest $\Rm$ (for the latter, see below).

\begin{figure}[!htp]

\caption{\label{cap:currentsheet} Like
Fig.~\ref{cap:Flow-Patterns-of}, but for $\Rm=6400$, $Re=400$.
Symmetry about $z=0$ has not been enforced; the jet forms
spontaneously at $z\approx -20$, but the whole pattern has been
shifted vertically to ease comparison with
Fig.~\ref{cap:Flow-Patterns-of}.}
\epsscale{1.00}
\plotone{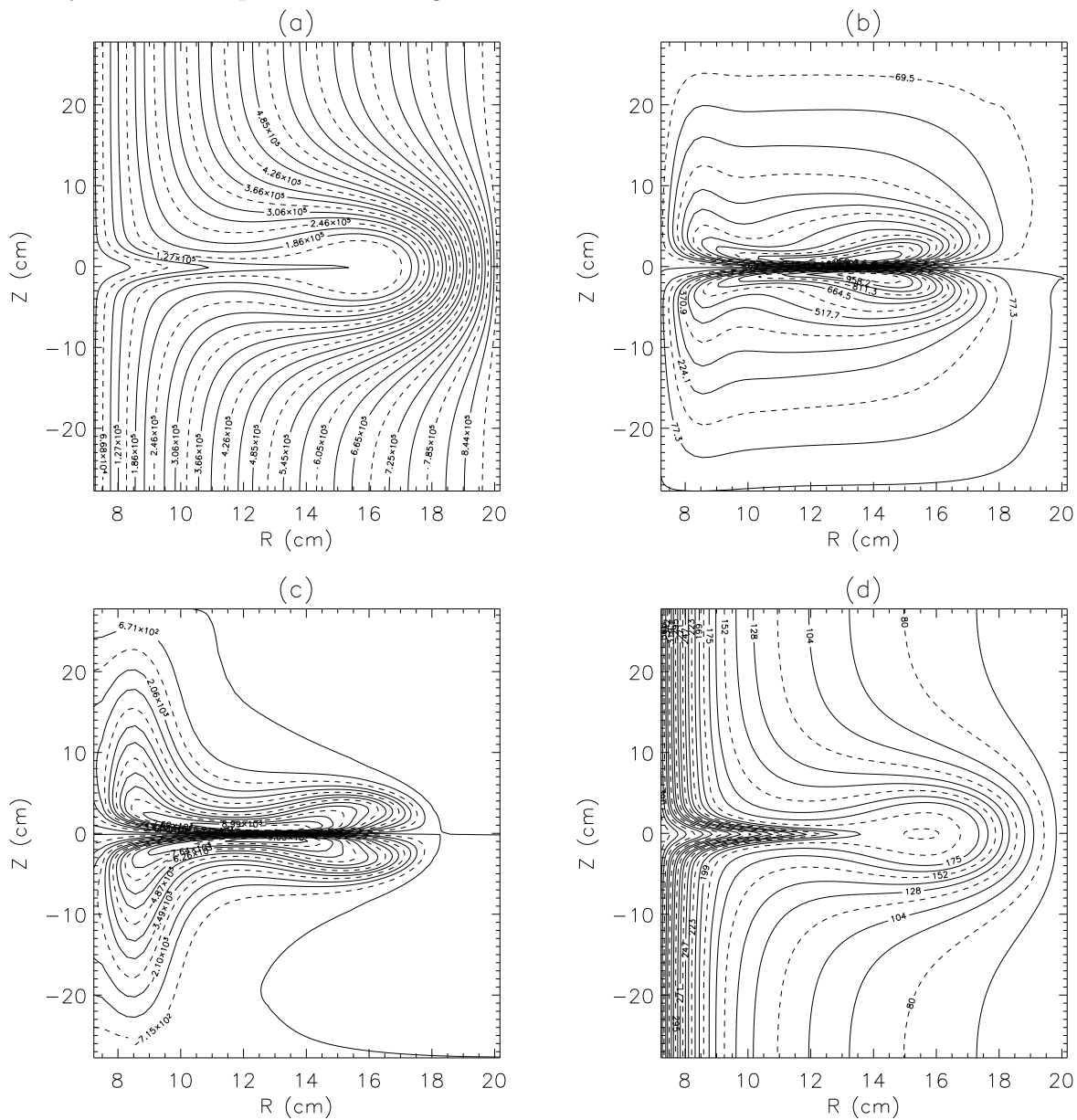}
\end{figure}

\begin{figure}[!htp]

\caption{\label{fit} Maximum radial current in the current sheet (left panel) and
maximum of toroidal magnetic field (right panel) \emph{vs.} magnetic Reynolds number,
for $Re=400$.  Powerlaw fits are shown as dashed lines with slopes $0.46$ [left panel]
and $0.18$ [right panel, eq~(\ref{btfit})].}
\epsscale{1.00}
\plotone{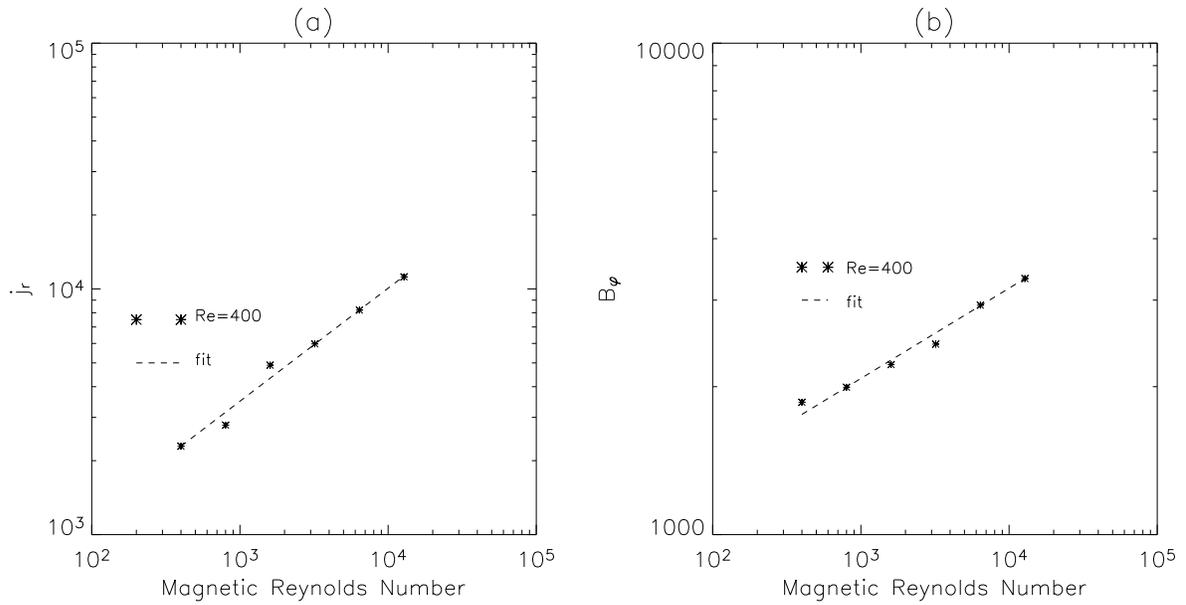}

\end{figure}

In steady state, the toroidal component of the electric field vanishes,
$E_\varphi=0$, because the flux through any circuit around the axis is
constant.  Consequently,
\begin{equation}\label{noEphi}
[\Phi,\Psi]\equiv\frac{\partial\Phi}{\partial r}\frac{\partial\Psi}{\partial z}
-\frac{\partial\Phi}{\partial z}\frac{\partial\Psi}{\partial r}=
\eta r\left(\frac{\partial^2}{\partial z^2}+\frac{\partial^2}{\partial r^2}
-\frac{1}{r}\frac{\partial}{\partial r}\right)\Psi\equiv \eta r\Delta_*\Psi,
\end{equation}
The evidence from our simulations is that the peak values of $\Phi$
and $\Psi$ scale as $\eta^{1/2}$ and $\eta^0$, respectively, in the
nonrestive limit $\eta\to0$, $\Rm\to\infty$.  The radial velocity
$V_r=r^{-1}\partial\Phi/\partial z$ also scales as $\eta^{1/2}$.  In
order that the two sides of eq.~(\ref{noEphi}) balance, at least one
of the derivatives of $\Psi$ must become singular in the limit
$\eta\to0$.  This appears to be the case.  In fact, a comparison of
the flux contours in Figures \ref{cap:Flow-Patterns-of}(a) and
\ref{cap:currentsheet}(a) suggests that a current sheet develops at
the center of the jet.  This is more obvious in the horizontal
components of current density, $J_r$ and $J_\varphi$, whose peak
values we find to scale as $\propto \eta^{-0.46} \approx \Rm^{1/2}$
(Figure~\ref{fit}) and the maximum toroidal magnetic field near the
current sheet scales as
\begin{equation}\label{btfit}
B_{\varphi} \propto \Rm^{0.18} \approx \Rm^{1/6}
\end{equation}
From these scalings one
infers that the width of the current sheet scales as $\eta^{1/3}$.
On the other hand, the region defined by $|B_r|, |B_\varphi| > |B_z|$
appears to have a width $\propto\eta^0$, like that of the
velocity jet.  We call this
the magnetic ``finger'' because of its form in Fig.~\ref{cap:currentsheet}.

It is interesting to check whether these scalings are consistent with
the observation that the total torque (radial angular-momentum flux)
appears to be asymptotically independent of the resistivity.  As
$\eta\to0$, the advective torque $\propto\int V_r V_\varphi dz$ tends
to zero since $V_r\propto\eta^{1/2}$ and $V_\varphi$ is presumably
bounded by $\sim r\Omega_1$.  The viscous contribution is always
dominant near the cylinders but is reduced compared to the initial
state at intermediate radii by the reduction in the
vertically-averaged radial shear (Fig.~\ref{cap:Z-Average-Torques-versus}).
Since the total torque is larger in the final than in the initial state, 
a significant fraction of it must be magnetic, and this fraction
should be approximately independent of $\eta$ at sufficiently small $\eta$.
If $B_r\sim B_\varphi\propto\eta^x$ within a vertical layer of width
$\Delta z\sim\eta^y$, the torque $\propto\int B_r B_\varphi dz\propto\eta^{2x+y}$.
Thus we expect $y\approx-2x$.  In agreement with this, we have found that
$x\approx-1/2$ and $y\approx1/3$ in the current sheet, while in the
finger, $x\approx y\approx 0$.

One notices in Fig.~\ref{cap:currentsheet}(a)\&(d) that the angular
velocity is approximately constant along field
lines---$\Omega=\Omega(\Psi)$---as required by Ferraro's Law when the
flow is predominantly toroidal and the resitivity small.  There must
therefore be an outward centrifugal force along the lines in the
magnetic finger, which in combination with the reconnection layer,
presumably drives the residual radial outflow.
Viscosity continues to be essential even as $\eta\to 0$ because it is
then the only mechanism for communicating angular momentum between
field lines, and between the fluid and the cylinders; the distortion of
the field enhances viscous transport by bringing into closer proximity
lines with different angular velocity.

To summarize, in the highly conducting limit $\Rm\to\infty$,
$Re=$constant, there appear to be at least three main regions of the
flow: (I) an ``external'' or ``reflux'' region in which the magnetic
field is predominantly axial and the velocity predominantly toroidal,
but with a small ($\propto\eta^{1/2}$) radial inflow; (II) a ``jet''
or ``finger'' of smaller but constant vertical width in which the
fields are mainly horizontal and there is a more rapid but still
$O(\eta^{1/2})$
flow along field lines; (III) a resistive layer or current sheet at the
center of the jet whose width decreases as $\eta^{1/3}$, across which
the horizontal fields change sign.

\subsection{Simulations at small magnetic Prandtl number}

In the ongoing Princeton MRI experiment, the experiment material,
liquid gallium, has kinematic viscosity
$\nu\approx3\times10^{-3}\;\unit{cm}^{2}\unit{s}^{-1}$ and resistivity
$\eta\approx2\times10^{3}\;\unit{cm}^{2}\unit{s}^{-1}$.  The typical
dimensionless parameters are $\Rm\approx 20$ and 
$Re\approx10^{7}$ at the dimensions and rotation
speeds cited above. The magnetic Prandtl number
$Pr\equiv\Rm/Re\approx10^{-6}$ is very small.  Reliable
simulations with
Reynolds number as high as $10^{7}$ are beyond any present-day
computer, and small $Pr$ presents additional challenges
for some codes.  

\begin{figure}[!htp]

\caption{\label{cap:Re25600_contours} Like Fig.~\ref{cap:Flow-Patterns-of},
but for $Re=25600$, $\Rm=20$.  The flow is unsteady but closely resembles
steady flows at lower $Re$ for this $\Rm$.}
\epsscale{1.00}
\plotone{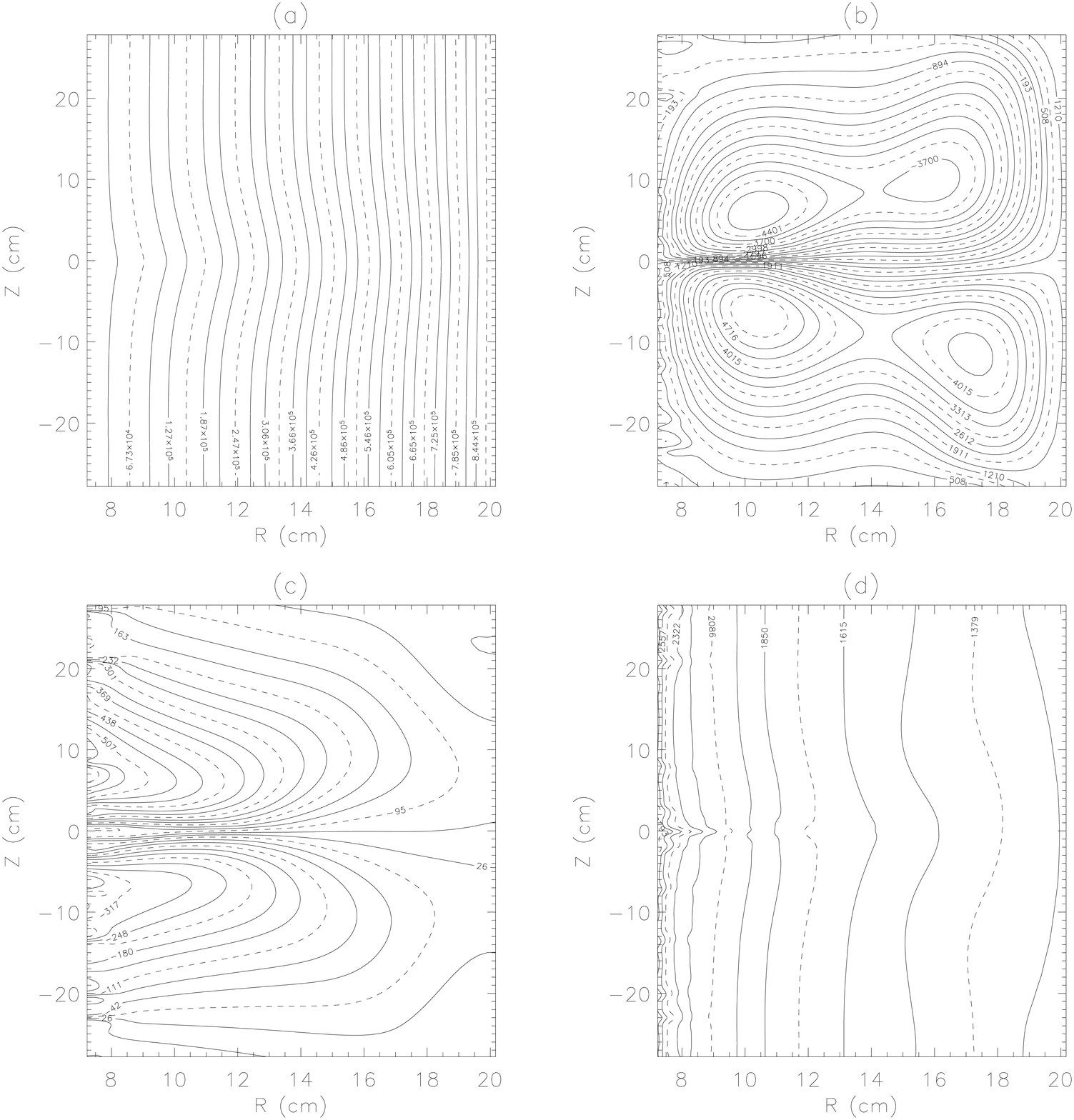}

\end{figure}

Although our boundary conditions are not those of the experiment, we
have carried out simulations at $\Rm=20$ and much higher $Re$ in order
to explore the changes in the flow due to these parameters alone.  A simulation
for $Re=25600$ is shown in Figures
\ref{cap:Re25600_contours} \& \ref{cap:Re25600_torques}.  All though this
is still considerably more viscous than the experimental flow, it is clearly
unsteady, like all of our simulations at $Re\gtrsim 3000$.
A narrow jet can still be observed in the poloidal velocities,
but the poloidal field is only weakly perturbed at this low
$\Rm$: $B_{\varphi,\max}\approx 0.1 B_z$.

Since the Reynolds number of the experiment
is much larger than
that of our simulations, we can estimate the experimental torques
only by extrapolation.
Extrapolating according to eq.~(\ref{torque_scaling})
from the highest-$Re$ simulation in Table~\ref{tab:torques}, one
would estimate $\Delta\Gamma/\Gamma_{\rm initial}\sim 35$ at $Re\sim 10^7$.
There are, however, reasons for caution in accepting this estimate.
On the one hand, the experimental flow may be three-dimensional and
turbulent, which might result in an even higher torque in the final state.
On the other hand, the viscous torque in the initial state is likely
to be higher than in these simulations
because of residual Ekman circulation driven by the split endcaps.
Nevertheless, we expect an easily measurable torque increase in the
MRI-unstable regime.

\section{Conclusions}\label{sec:5}

\begin{figure}[!htp]
\caption{Total toroidal magnetic energy \emph{vs.} time at $Re=400$.}
\label{approach}
\epsscale{0.6}
\plotone{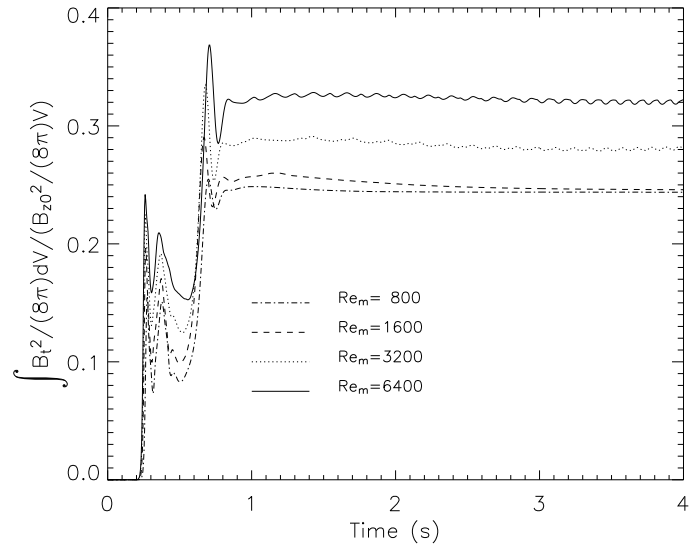}

\end{figure}

\begin{figure}[!htp]

\caption{\label{cap:Re25600_torques} The
$z$-averaged torques as in
Fig.~\ref{cap:Z-Average-Torques-versus},
but for the state shown in Fig.~\ref{cap:Re25600_contours} 
($Re=25600$, $\Rm=20$).  The radial variation of the total torque, though
slight, testifies to the unsteadiness of the flow.}
\epsscale{1.00}
\plotone{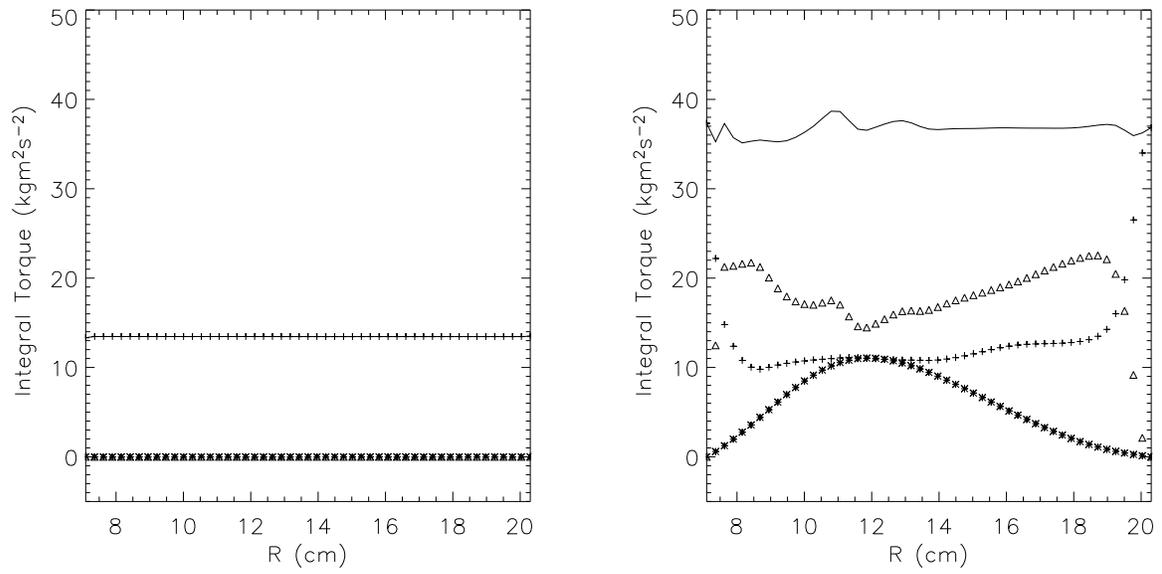}

\end{figure}

In this paper, we have simulated the linear and nonlinear development
of magnetorotational instability in a nonideal magnetohydrodynamic
Taylor-Couette flow.  The geometry mimics an experiment in preparation
except in the vertical boundary conditions, which in these simulations
are periodic in the vertical (axial) direction and perfectly
conducting at the cylinders; these
simplifications allow direct contact with previous linear studies.  We
have also restricted our study to smaller fluid Reynolds number ($Re$), and
extended it to larger magnetic Reynolds number ($\Rm$), than in the
experiment.  We find that the time-explicit compressible MHD code
ZEUS-2D, which is widely used by astrophysicists for supersonic ideal
flows with free boundaries, can be adapted and applied successfully to
Couette systems.  MRI grows from small amplitudes at rates in good agreement
with linear analyses under the same boundary conditions.
Concerning the nonlinear final state that results from saturation of
MRI, we draw the following conclusions:
\begin{itemize}
\item Differential rotation is reduced except near
boundaries, as predicted by \citet{kj05}.
\item A steady poloidal circulation consisting of a narrow outflow
(jet) and broad inflow is established.  The width of the jet is
almost independent of resistivity, but it does decrease with
increasing $Re$.  The radial speed of the jet $\propto\Rm^{-1/2}$.
\item There is a reconnection layer within the jet whose width
appears to decrease $\propto \Rm^{-1/3}$.
\item The vertically integrated radial angular momentum flux
depends upon viscosity but hardly upon resistivity, at least
at higher $\Rm$ [eq.~(\ref{torque_scaling})].
\item The final state is steady and laminar at $Re,\Rm\lesssim 10^3$
but unsteady at larger values of either parameter
(Figs.~\ref{approach} \& \ref{cap:Re25600_torques}.)
\item the final state contains horizontal fields
comparable to the initial axial field for $\Rm\gtrsim 400$,
and about a tenth as large for experimentally more realistic
values, $\Rm\approx 20$.
\end{itemize}

We emphasize that these conclusions are based on axisymmetric
simulations restricted to $10^2\lesssim Re,\Rm\lesssim 10^{4.4}$, and that
the boundary conditions are not realistic.  This paper is intended
as a preliminary exploration of MRI in the idealized Taylor-Couette
geometry that has dominated previous linear analyses.  We have not
attempted to model many of the complexities of a realistic flow.
In future papers, we will study vertical boundary conditions closer to
those of the planned experiment; work in progress indicates that these
may significantly modify the flow.

\begin{acknowledgements}

The authors would like to thank James Stone for the advice on the ZEUS
code. This work was supported by the US Department of Energy, NASA
under grant ATP03-0084-0106 and APRA04-0000-0152 and also by the
National Science Foundation under grant AST-0205903.

\end{acknowledgements}



\clearpage

%


%

\clearpage
\begin{table}[!htp]

\caption{\label{cap:Gauss-Magnetic-Diffusion}Magnetic Diffusion Test}

\begin{tabular}{|c|c|c|c|c|}
\hline 
$\Rm$&
Resolution &
Decay Rate [$s^{-1}$]&
Exact Rate&
Error (\%)\tabularnewline
\hline
\hline 
400&
100x100&
382.52642&
392.26048&
2.482\tabularnewline
\hline 
400&
50x50&
352.76963&
391.87454&
9.979\tabularnewline
\hline 
100&
100x100&
1533.6460&
1569.0419&
2.256\tabularnewline
\hline 
100&
50x50&
1420.4078&
1567.4982&
9.384\tabularnewline
\hline
\end{tabular}
\end{table}

\begin{table}[!htp]

\caption{\label{cap:Comparison-with-Global} Growth rates from
semianalytic linear analysis \emph{vs.} simulation.}
\begin{tabular}{|c|c|c|c|c|}
\hline 
$\Rm$&
$Re$&
$~n~$ &
Prediction $[\unit{s}^{-1}]$&
Simulation $[\unit{s}^{-1}]$ \tabularnewline
\hline\hline
&
&
1&
41.67&
\tabularnewline
&
&
2&
72.71&
\tabularnewline
400&
400&
3&
77.69{*}&
77.66{*}{*}\tabularnewline
&
&
4&
56.88&
\tabularnewline
&
&
5&
0.283&
\tabularnewline
\hline 
&
&
1&
23.31&
\tabularnewline
20&
$\infty$ &
2&
32.43{*}&
30.83{*}{*}\tabularnewline
&
&
3&
23.73&
\tabularnewline
&
&
4&
6.905&
\tabularnewline
\hline 
\end{tabular}
\end{table}

\begin{table}[!htp]

\caption{\label{tab:torques}Increase of total torque
versus $Re$ and $\Rm$.}

\begin{tabular}{|c|c|c|c|c|c|}
\hline
$\Rm$&
$Re$&
Resolution&
$\Gamma_{\rm initial}$ &$\Gamma_{\rm final}$&$\Delta\Gamma/
\Gamma_{\rm initial}$ \tabularnewline
& &$N_{z} \times N_{r}$ &$[\unit{kg\,m^{2}\,s^{-2}}]$ & $[\unit{kg\,m^{2}\,s^{-2}}]$ &\\
\hline
\hline
10&
400&
200$\times$50&
8.60e2&
8.60e2&
0.00\tabularnewline
\hline
20&
400&
200$\times$50&
8.60e2&
9.08e2&
0.06\tabularnewline
\hline
50&
400&
200$\times$50&
8.60e2&
1.12e3&
0.30\tabularnewline
\hline
100&
400&
200$\times$50&
8.60e2&
1.35e3&
0.57\tabularnewline
\hline
200&
400&
200$\times$50&
8.60e2&
1.50e3&
0.74\tabularnewline
\hline
400&
400&
200$\times$50&
8.60e2&
1.57e3&
0.83\tabularnewline
\hline
800&
400&
200$\times$50&
8.60e2&
1.57e3&
0.83\tabularnewline
\hline
1600&
400&
200$\times$50&
8.60e2&
1.67e3&
0.94\tabularnewline
\hline
3200&
400&
200$\times$50&
8.60e2&
1.65e3&
0.92\tabularnewline
\hline
6400&
400&
200$\times$50&
8.60e2&
1.62e3&
0.88\tabularnewline
\hline
12800&
400&
228$\times$50&
8.60e2&
1.62e3&
0.88\tabularnewline
\hline
\hline
400&
100&
200$\times$50&
3.44e3&
4.45e3&
0.44\tabularnewline
\hline
400&
200&
200$\times$50&
1.72e3&
2.58e3&
0.50\tabularnewline
\hline
400&
400&
200$\times$50&
8.60e2&
1.57e3&
0.83\tabularnewline
\hline
400&
800&
200$\times$50&
4.30e2&
9.70e2&
1.26\tabularnewline
\hline
  400&
1600&
200$\times$50&
2.15e2&
6.20e2&
1.88\tabularnewline
\hline
  400&
3200&
200$\times$50&
1.08e2&
3.90e2&
2.63\tabularnewline
\hline
  400&
6400&
200$\times$50&
5.38e1&
2.46e2&
3.58\tabularnewline
\hline
  400&
12800&
228$\times$58&
2.69e1&
1.55e2&
4.77\tabularnewline
\hline
\hline
20&
100&
200$\times$50&
3.44e3&
3.44e3&
0.00\tabularnewline
\hline
20&
200&
200$\times$50&
1.72e3&
1.72e3&
0.00\tabularnewline
\hline
20&
400&
200$\times$50&
8.60e2&
8.95e2&
0.04\tabularnewline
\hline
20&
800&
200$\times$50&
4.30e2&
4.95e2&
0.15\tabularnewline
\hline
20&
1600&
200$\times$50&
2.15e2&
2.76e2&
0.28\tabularnewline
\hline
20&
3200&
200$\times$50&
1.08e2&
1.57e2&
0.45\tabularnewline
\hline
20&
6400&
200$\times$50&
5.38e1&
9.35e1&
0.74\tabularnewline
\hline
20&
12800&
228$\times$50&
2.69e1&
5.75e1&
1.14\tabularnewline
\hline
20&
25600&
320$\times$50&
1.34e1&
3.70e1&
1.75\tabularnewline
\hline

\end{tabular}
\end{table}

\end{document}